\documentstyle[11pt,epsf,aaspp4]{article}
\lefthead{F. Frontera et~al.}
\righthead{GRB981226 with BeppoSAX}

\def\S{{\em BeppoSAX\/}}   
\def\etal{{\it et al. }}

\def\ergcm{\mbox{ erg cm$^{-2}$}}
\def\apj{{\it Astrophys. J. }}
\def\apjs{{\it Astrophys. J. Suppl. Ser. }}

\def\aa{{\it Astron. Astrophys. }}
\def\aas{{\it Astron. Astrophys. Suppl. Ser. }}

\def\ergcms{\mbox{ erg cm$^{-2}$ s$^{-1}$}}
\makeatletter
\def\@cite#1#2{(#1\if@tempswa , #2\fi)}
\def\preprint{preprint}   \newif\ifPreprintMode
\ifx\preprint\revtex@genre\PreprintModetrue\else\PreprintModefalse\fi
\makeatother

\begin{document}

\title{Prompt and afterglow emission from the X-ray rich GRB981226 
observed with BeppoSAX}

\author{F. Frontera\altaffilmark{1,2},
L.A.~Antonelli\altaffilmark{3},
L.~Amati\altaffilmark{2},
E.~Montanari\altaffilmark{1},
E.~Costa\altaffilmark{4},
D.~Dal~Fiume\altaffilmark{2},
P.~Giommi\altaffilmark{6},
M.~Feroci\altaffilmark{4},
G.~Gennaro\altaffilmark{11},
J.~Heise\altaffilmark{5},
N.~Masetti\altaffilmark{2},
J.M.~Muller\altaffilmark{5,6},
L.~Nicastro\altaffilmark{7},
M.~Orlandini\altaffilmark{2},
E.~Palazzi\altaffilmark{2},
E.~Pian\altaffilmark{2},
L.~Piro\altaffilmark{4},
P.~Soffitta\altaffilmark{4},
S.~Stornelli\altaffilmark{11}
J.J.M.~in 't Zand\altaffilmark{5},
D.A.~ Frail\altaffilmark{8},
S.R.~Kulkarni\altaffilmark{9},
and M. Vietri\altaffilmark{10}
}

\altaffilmark{1}{Dipartimento di Fisica, Universit\`a di Ferrara, Via Paradiso
 12, 44100 Ferrara, Italy}

\altaffilmark{2}{Istituto Tecnologie e Studio Radiazioni Extraterrestri, 
CNR, Via Gobetti 101, 40129 Bologna, Italy}

\altaffilmark{3}{Osservatorio Astronomico di Roma, Via Frascati, 33, 00040
Monteporzio Catone (RM), Italy}

\altaffilmark{4}{Istituto Astrofisica Spaziale, C.N.R., Via Fosso del 
Cavaliere, 00133 Roma, Italy}

\altaffilmark{5}{Space Research Organization in the Netherlands,
 Sorbonnelaan 2, 3584 CA Utrecht, The Netherlands}

\altaffilmark{6}{\S\ Scientific Data Center, Via Corcolle 19, 00131 Roma,
 Italy} 

\altaffilmark{7}{Istituto Fisica Cosmica e Applicazioni all'Informatica, 
C.N.R., Via U. La Malfa 153, 90146 Palermo, Italy}

\altaffilmark{8}{National Radio Astronomy Observatory, P.O. Box O, Socorro,
NM 87801, USA}

\altaffilmark{9}{Palomar Observatory 105-24, Caltech, Pasadena, CA 91125, USA}

\altaffilmark{10}{Dipartimento di Fisica, Universit\`a Roma Tre, Via della 
Vasca Navale, 84, 00146 Roma, Italy}

\altaffilmark{11}{\S\ Operative Control Center, Via Corcolle 19, 00131 Roma,
 Italy}

\begin{abstract}
We discuss observations  of the prompt X-- and $\gamma$--ray emission and X--ray afterglow
 from GRB981226. This event has the weakest gamma-ray peak flux
detected with the \S\ Gamma-Ray Burst Monitor. It shows an isolated X-ray precursor
and the highest X-ray to gamma-ray fluence ratio measured thus far with the \S\
Wide Field Cameras. The event was followed up 
with the \S\ Narrow Field Instruments, and the X-ray afterglow was detected up to 10 keV. 
The afterglow flux is observed to rise from a level below the sensitivity of the
MECS/LECS telescopes up to a peak flux of (5$\pm 1) \times 10^{-13}\, 
\ergcms$ in the 2-10~keV energy band. This rise is followed by a  decline according to a 
power law with index of 1.31$^{+0.44}_{-0.39}$. We discuss these results 
in the light  of the current GRB models.
\end{abstract}

\keywords{gamma rays: bursts --- gamma rays: observations --- X--rays:
general ---shock waves}

\section{Introduction}
Follow-up observations of arcminute positions of Gamma-Ray Bursts (GRBs) 
provided by provided by the Wide Field Cameras on the \S\ satellite \cite{Jager97,Boella97a} 
have shown that in most
cases a fading X-ray counterpart identified as the GRB X-ray afterglow is
detected \cite{Frontera98}. The fading law  is generally  a smooth power law 
(e.g., Frontera et~al. 1999\nocite{Frontera99}) except in two cases: GRB970508 
\cite{Piro98}, in which a late-time outburst of about 10$^5$~s duration 
started about 6$\times 10^4$s after the main event, and GRB970828 
\cite{Yoshida99}, in which a peak structure of 4000~s duration appeared 
1.25 $\times 10^5$~s after the main event. Generally the 
afterglow light curves extrapolated back to the time of the bursts, are in 
agreement with the tail of the GRB time profiles 
\cite{Costa97,Piro98,Frontera99}. This fact is
considered as evidence that late afterglow emission and tail of the prompt 
GRB emission have the same origin \cite{Frontera99}. In the fireball model 
scenario (see, e.g., the recent review by Piran 1999 \nocite{Piran99}), 
this means that both the tail of the prompt emission 
and the late afterglow emission can be due to an external shock produced 
by the interaction of a relativistically expanding fireball  
with the Interstellar Medium (ISM).  
\\
Of the promptly localized GRBs  80\% show X-ray afterglow,
about 50\% exhibit also optical emission, and 30\% 
show radio emission. Radio emission is generally accompanied by optical 
emission, except in two cases: GRB990506 \cite{Taylor99}
from which also X-ray afterglow emission was observed 
with the RXTE/PCA experiment (Bacodine trigger 7549, Hurley 1999 \nocite{Hurley99}), 
and GRB981226. For this event, in spite of several attempts to find the 
optical counterpart 
\cite{Galama98,Rhoads98,Bloom98,Schaefer98,Castro-Tirado98,Wozniak98,Wozniak98,Lindgren99}
none was identified. The best upper limit (R$\sim$23~mag) to the optical flux of the 
GRB counterpart was reported by Lindgren \etal (1999) \nocite{Lindgren99}. 
Frail \etal (1999a) \nocite{Frail99a} reported the detection of a 
radio counterpart (VLA 232937.2--235553).  
The radio source peaked to 173$\pm27$~$\mu$Jy at 8.46~GHz at about 10 days after 
the burst. This time delay is tipical of  
all previously studied radio afterglows. However, the source declined relatively fast, 
following a power law decay ($\propto t^{-\delta_R}$) with 
$\delta_R \, = \, 2.0 \pm 0.4$ \cite{Frail99b}. Two interpretations of this
rapid decay have been proposed by Frail \etal (1999b) \nocite{Frail99b}: it is 
either the consequence of a jetted GRB or the result of a fireball shock in an ambient 
medium with variable density.
An optical galaxy (R~=~24.85 mag), consistent with the radio transient
position, has been proposed as the host galaxy of GRB981226 \cite{Frail99b}.

GRB981226 was also followed up with the \S\ Narrow Field Instruments. A transient X-ray 
source was observed, which was proposed as the X-ray afterglow of GRB 981226 
\cite{Frontera98}.
Here we report the properties of this afterglow emission  along 
with those of the prompt X-- and gamma--ray emission. We will discuss these 
properties in the light of the current models of GRBs.

\section{Observations} 
\label{obs}
GRB981226 was detected with the \S\ Gamma--Ray Burst Monitor (GRBM, 
40--700~keV; Frontera et~al. 1997, Amati et~al. 1997\nocite{Frontera97,Amati97}) and 
WFC unit 1 (1.5--26.1~keV; Jager et~al. 1997\nocite{Jager97}) on December 26 starting 
at 09:47:25 UT \cite{Diciolo98}. Its 
position was determined with an error radius of $6'$ (99\% confidence level) 
and was centered at $\alpha_{2000} = 23^{\rm h}29^{\rm m}40^{\rm s}$, $\delta_{2000} =
-23^\circ55'30''$. A precursor is detected only in the WFC on 09:44:20 UT.
 
About eleven hours after the burst, the Narrow Field Instruments 
on-board \S\ were pointed  at the burst location for a first
target of opportunity (TOO1) observation, from December 26.8785 UT to December
28.2986~UT. A new X--ray source was detected
\cite{Frontera98} in the GRB error box with the Low Energy (LECS, 0.1-10~keV, 
Parmar et~al. 1997 \nocite{Parmar97}) and Medium Energy (MECS, 2--10~keV,
Boella et~al. 1997b\nocite{Boella97b}) Concentrators/Spectrometers. The net exposure 
time on the source for MECS
and LECS was 58.4~ks and 25.6~ks, respectively. The same field was
again observed about 7 days after the main event (TOO2) from 1999
January 2.7604~UT to
January 3.5590~UT (total net exposure time of 25.1~ks for MECS and 8.0~ks for
LECS).  During this observation, the source was no more detected.

Data available from GRBM  include two 1~s ratemeters in two energy
channels (40--700~keV and $>$100~keV), 128~s  count spectra (40--700~keV,
225 channels) and high time resolution data (up to 0.5~ms) in the
40--700~keV energy band.
%The energy resolution of the GRBM unit 1, co-aligned 
%with the WFC No. 1, is 20\% at 280 keV \cite{Amati97}. 
WFCs (energy resolution $\approx$ 20\% at 6~keV) were operated in normal
mode with 31 channels in 1.5--26~keV and 0.5~ms time resolution 
\cite{Jager97}. 
The burst direction was offset by 7$^\circ$ with respect to
the WFC axis. With this offset, the
effective area exposed to the GRB was
$\approx$~420~cm$^2$ in the 40-700~keV band and  91~cm$^2$
in the 2--26~keV energy band.
The background  in the WFC and GRBM energy bands  was fairly stable
during the burst, with a slight increase with time in the $>$100~keV
channel (about 2\% in 350~s). The GRBM background was estimated by linear 
interpolation using the 250~s count rate data  before and after the burst.
The WFC spectra were extracted through the Iterative Removal Of Sources procedure
(IROS \footnote{WFC software version 105.108}, e.g. Jager et al. 1997 \nocite{Jager97})
which implicitly subtracts the contribution of the background and of other point sources 
in the field of view.
\\
The MECS source count rates and
spectra for TOO1 were extracted, using the XSELECT package, from a $\sim 3'$ radius region
around the source centroid, while the background level was estimated
from an annulus centered on the source with inner and outer radii of 4 and 
8.5 arcmin, respectively. The spectra from MECS 2 and 3 were equalized and co-added.
Given the much lower exposure time, the source was much less visible in the LECS. We used
for the source extraction from the LECS image the XIMAGE package \cite{Giommi91}, 
that permits a more
refined choice of the background region. The source counts  were extracted 
from a square box centered on the source centroid consistent with the MECS centroid 
position  and with side of 3.5 arcmin, while the corresponding background was 
extracted from a square annulus of
inner side 3.5 arcmin and outer side 12 arcmin, centered on the source.
The uncertainties will be given as single parameter errors at 90\% confidence level.

\section{Results}

\subsection{Prompt emission}
\label{prompt}
Figure~1 shows the measured time profiles of GRB981226 in three energy channels after the 
background subtraction.
In the $\gamma$--ray band (40--700 keV; fig.~1, middle panel), the GRB shows a single peak 
of about 5~s duration. Some marginal evidence of the peak appears in the high energy range 
($>$100 keV; fig.~1, bottom panel).
In the X--ray energy band (2--26~keV; fig~1, top panel), the prompt emission starts about 180~s 
earlier with a precursor-like event of about 50~s duration. The X--ray main event
exhibits two peaks, the first of which is coincident with that detected by the GRBM.
The total duration of X-ray main event is  about 80~s. From the WFC images both precursor 
and main event are consistent with the same direction in the sky.

The spectral evolution of  precursor and main event was studied
by subdividing  the GRB time profile into five temporal slices and 
performing an analysis on the average spectrum of each
slice (see Fig.~1). 
We fit the spectra  with a power law (N(E)$\propto \ E^{\alpha}$) and  a smoothly 
broken power law 
\cite{Band93}, both  photoelectrically  absorbed by a neutral hydrogen column
density N$_H$ \cite{Morrison83}. The count statistics 
do not permit to constrain N$_H$, that was thus fixed to the 
Galactic value along the GRB direction (1.8 $ \times 10^{20}\,{\rm cm^{-2}}$).  
In Table 1 we show the results. Both laws fit the data. The Band law permits
to determine the value of the peak energy E$_p$ of the logarithmic power per photon energy
decade (the $\nu F_\nu$ spectrum). For the time slices A, C, D, where only upper
limits to the gamma-ray flux were available, the upper limits of E$_p$  were derived
by freezing the value of the high energy index $\beta$ to $-2$.
In Fig.~2 we show the $\nu F(\nu)$ spectra of the chosen temporal slices. 
The continuous line is the best fit of the Band law for slice B and of the power
law for the other slices.
A spectral softening is observed from slice B (onset of the main 
event) to the following slices. Also the precursor spectrum appears softer 
than that in the slice B. The spectral evolution is better shown by the behavior
of the peak energy E$_p$ (see Table~1): it is low of our energy passband during the
precurson, it achieves a value of about 60 keV at the onset of the main event 
and then goes down below our passband at the end of the burst.
\\
The $\gamma$--ray (40--700~keV) fluence of the burst is S$_\gamma\,=\, 
(4\pm 1)\times 10^{-7}$ \ergcm, while the corresponding value found
in the 2--10 keV band is S$_X\,=\,(5.7\pm 1.0)\times 10^{-7}$~\ergcm, with
a ratio S$_X$/S$_\gamma$ \,=\,(1.4$\pm$0.4).
%In the 50--300~keV range the fluence of GRB981226 is 2.8$\times 10^{-7} \ergcm$.
The $\gamma$--ray peak flux, derived from the 1~s ratemeters, is P$_\gamma\,=\, 
0.33$$\pm$0.13~photons/cm$^2$~s corresponding to 
$(6.5\pm2.6)\times 10^{-8}$ \ergcms, while the
corresponding 2--10 keV peak flux is  P$_X\,=\,2.7$$\pm$0.3~photons/cm$^2$~s,
corresponding to $(1.7\pm0.2)\times 10^{-8} \ergcms$.
The peak flux in gamma-rays is the lowest observed thus far with \S\ .   
\\
\subsection{Afterglow emission}
\label{after}
During TOO1, a previously unknown X-ray source, 1SAX J2329.6-2356,  was detected 
in the MECS, almost at the center of the GRB error box, at celestial coordinates
$\alpha_{2000} = 23^{\rm h}29^{\rm m}36.1^{\rm s}$,
$\delta_{2000} = -23^\circ55'58.3''$, with an error radius of 1 arcmin \cite{Frontera98}.
The MECS image obtained in the first part of TOO1 (exposure time of 26950~s), when
the source was stronger, is shown in fig.\ref{fig:mecs_image_1_2}, left.
The source is also visible in the corresponding  0.1--2~keV LECS image (6880 exposure 
time).   
%The MECS and LECS source exposure times during TOO1 were  26950~s and 6880~s, 
%respectively. 
The source was not detected in the TOO2, when the MECS exposure time was 25080~s 
(see fig.~3, right).
\\
Other count excesses, compatible with very weak celestial 
sources, are present in the TOO1 image. They are likely field sources, 
the number of which
is in agreement with the log~N-log~S distribution of the 5-10~keV X-ray 
sources found by Fiore  et al. (1999b) \nocite{Fiore99b} with \S\ . Given its 
transient behavior, 1SAX J2329.6-2356 is likely the X-ray aftergow 
of GRB981226. From the above log~N-log~S distribution, the chance probability for 
its coincidence with a background source is about 5$\times 10^{-3}$. 

\subsubsection{Spectrum}
\label{spectrum}
We derived the average 0.1-10~keV count spectrum of 1SAX J2329.6-2356 over the 
first 15~hrs of TOO1, when the source was brightest. We fit it both with a power
law (N(E)$\propto \ E^{\alpha}$) and a blackbody, photoelectrically absorbed by the Galactic 
column density along the GRB direction (see Section \ref{prompt}). In the fits a
normalization of a factor 0.8 was applied to the LECS spectra following the
cross-calibration tests between the LECS and MECS \cite{Fiore99a}. 
Both laws are acceptable descriptions of the data: $\chi^2_\nu\,=\,0.5$ (5 degrees of 
freedom, dof) 
for a power-law and $\chi^2_\nu\,=\,1.3$ (5 dof) for a blackbody. However, 
in the case of the
blackbody, for energies above 5~keV, the best fit curve is constantly below 
the measured bins. This suggests that the power law provides a better description 
of the data.
The best fit power-law index is $\alpha \,= \, -1.92 \pm 0.47 $.
We do not find evidence of a spectral evolution of the emission: the time
behavior of the C(4--10~keV)/ C(1.4--4~keV) hardness ratio is statistically consistent 
with a constant. 
\\
No evidence of the source is found in the 15--300~keV energy range: the 
\S\ PDS instrument \cite{Frontera97} during TOO1 does not show  any statistically 
significant count excess over the background level. Assuming the above power-law 
index, the  2~$\sigma$ upper limit in the 15--60~keV energy band is  
$5.5\times 10^{-12} \ergcms$, which is  
a factor $\sim$30 higher than the extrapolated flux from the LECS+MECS spectrum.
\\
As above mentioned, no source excess is apparent in the MECS~+~LECS data
during TOO2. Assuming the best fit power law index obtained from the
TOO1 spectrum and the Galactic column density,
we derived the following 2~$\sigma$ upper limits to the source flux during
TOO2: $1.3\times 10^{-13} \ergcms$ and $8.2 \times 10^{-14} \ergcms$ in the 0.1--2~keV
and 2--10~keV ranges, respectively.

\subsubsection{Light curve}
\label{lc}

The 2--10~keV MECS light  curve of the afterglow  in bins of 
7000~s elapsed time  is shown in fig.~\ref{fig:lc} (top).
Its main feature is the weakness of the source in the first 7000~s bin, where its flux
is below the MECS sensitivity limit (2$\sigma$ upper limit of $1.5 \times 10^{-13} 
\ergcms$ in the 2-10~keV energy band).
Checks were done to verify whether  this non detection could be due to attitude 
malfunctions of the satellite. However we found that the source was correctly 
pointed since the beginning of the NFI TOO1. Afterwards the source flux increases by
more than a factor 3 in about 10000~s ((5$\pm 1)\times 10^{-13} \ergcms$). After this 
peak flux the source starts fading.
The light curve is also reported in the bottom panel of fig.~\ref{fig:lc} along
with the WFC data points.  The later fading of 1SAX J2329.6-2356 is apparent, that
is well described by a power law, F(t)$\propto$~t$^{-\delta}$,  
with index $\delta \,= \, 1.31^{+0.44}_{-0.39}$. In fig.~\ref{fig:lc}, bottom 
we show  the best fit power law along with the slope uncertainty region (90\% 
confidence level).
From the best fit light curve, the 2--10~keV afterglow fluence
integrated over the time interval from the GRB end (80~s) to 10$^6$~s is
$S_a \, = \, (4.3\pm 3.8)\times 10^{-7} \ergcm$, with
a ratio between X--ray afterglow fluence and the prompt $\gamma$--ray 
fluence of 1.1$\pm$0.9, vs. a corresponding value of (1.4$\pm$0.4) for the 
prompt emission (see Sect. \ref{prompt}).

\section{Discussion}
GRB981226 shows the lowest gamma-ray peak flux among the bursts localized thus 
far with \S\ . In the log~N--log~P distribution of the GRBs observed with BATSE 
\cite{Paciesas99}, it is located near the faint end of the distribution.
The burst is marked by three peculiarities, two of which have been
seen for the first time:

\begin{enumerate}

\item
Of the bursts localized by BeppoSAX, this burst is the richest in the
X-ray band:  the X-ray to gamma-ray fluence ratio of $1.4\pm 0.4$ is
the highest of all SAX bursts \cite{Frontera99}. The peak energy
$E_p\sim 60$ keV is softer than that measured in other \S\  bursts.
\item
An isolated X-ray precursor occurs about 180 s before the main event (Figures
1--2). Onset of X-ray emission before the gamma-rays has been observed
in other GRBs \cite{Laros84,Murakami91,Zand99,Feroci99}, but only another isolated X-ray 
precursor, started about 25~s before the start of $\gamma$--rays, has been
reported thus far \cite{Laros84}.
\item
The X-ray afterglow light curve is peculiar. As can be seen from
Figure 4 (top), the afterglow emission is undetectable during first 
2 hours of the start of the NFI observations (2$\sigma$ upper
limit of $1.5 \times 10^{-13} \ergcms$ in 2-10~keV), after which it rises
rapidly to (5$\pm 1) \times 10^{-13} \ergcms$, then undergoes a decline in the 
typical power law fashion (index $\delta \,= \, 1.31^{+0.44}_{-0.39}$).
Thus at least during epoch 11--13 hrs the afterglow is undetectable.
\end{enumerate}

Item (2) is an uncommon feature of GRBs. Assuming the synchrotron shock
model (e.g., Piran 1999 \nocite{Piran99}), it implies a 
an initial fireball expansion Lorentz factor smaller than that found in other GRBs
(e.g., Frontera et~al. 1999\nocite{Frontera99}).
\\
Item (3) is the most interesting and mysterious aspect of this burst.
In all SAX observed bursts to date, the afterglow emission at X-ray
wavelengths begins more or less as soon as the main gamma-ray event
ends. Also in the case of GRB981226 we have evidence that the afterglow starts
during the main event. Indeed the peak flux of the X-ray prompt emission 
($\gtrsim 300$~$\mu$Jy) has the same order of magnitude as that
(173$\pm27$~$\mu$Jy at 8.46~GHz) observed in the radio band $\sim$10 days after the 
burst \cite{Frail99b}. On the contrary, the X-ray  flux measured by us about 13~hrs 
after the 
main event is about 4 orders of magnitude lower than the radio peak flux. Now
similar values of peak fluxes in X-rays and in the radio band  are 
expected in  quasi-adiabatic cooling shocks of relativistically expanding 
fireballs \cite{Sari98}.     
Thus a simplest conclusion is that the afterglow emission began immediately after 
the burst 
(phase E in fig.~4), rather than 13~hrs later, and then it was strongly reduced or
completely ceased, and eventually restarted 13 hours after the burst.
If we accept this explanation, then we must explain the physical cause for this
rather extended gap.
\\
The afterglow emission is attributed to the forward shock, i.e. shock
of the ambient gas particles swept up by the advancing blast wave.
A cessation of X-ray afterglow would require that there be no ambient
gas (or very reduced density) and the resurgence of the afterglow at
13 hr would then require increased density -- in short, a cavity
surrounding the explosion. The rapid decline of the X-ray and radio
afterglow require that the ambient density not be a constant but
decrease with the radius \cite{Chevalier99}.

Indeed, one expects such a circumburst medium around massive stars to
have a complicated geometry. For example, a  star which exploded as a
blue supergiant would have suffered two episodes of mass loss. During
the first phase, the red supergiant phase, the wind speed is low
leading to a rich circumstellar medium. During the next phase, the blue
supergiant phase, the fast blue supergiant wind sweeps up the
circumburst medium shaped by the red supergiant wind with the
net result of a low-density cavity surrounded by a dense
shell at the outskirts. 

If one accepts this explanation then we have found an additional evidence
linking GRBs to massive stars. In this scenario, we find a satisfying
explanation also for item (1) and (2). The X-ray richness is because the
blast wave picks up matter (baryons) as it tunnels through the envelope
of the massive star. The resulting large baryon content leads to lower
$E_p$ and hence more X-ray emission.  We attribute the percursor
emission to emission from shock breakout that is a natural consequence
of models in which GRBs arise from the death of massive stars \cite{MacFadyen99}.

The rapid decline in the radio and X-ray afterglow is because of
the radial density gradient in the circumburst medium. Finally, the
absence of an optical afterglow could well be due to extinction towards
the GRB. If this GRB arises from death of a massive star then most
likely the progenitor was in a dusty region and hence the extinction.

We tend to exclude that the X-ray increase of the afterglow  observed $\sim$13~hrs 
after the burst is a signature of a supernova emission. Given that the X-ray
flux is similar to that measured for the X-ray counterpart of SN1998bw \cite{Pian99}, one would
expect a moderate distance for GRB981226 (z$\lesssim$~0.01) and therefore a significant 
detection in the optical band also in the case of a large extinction.

\acknowledgements

This research is supported by the Italian Space Agency (ASI). 
We thank the \S\
mission director R.C.\ Butler and the teams of the \S\ Operative Control
Center and Scientific Data Center
for their efficient and enthusiastic support to the GRB alert program.

\clearpage

\begin{deluxetable}{cccccc}
%\small
%\footnotesize
%\scriptsize
\tablewidth{0pt}
\tablenum{1}
\tablecaption{Spectral evolution of GRB981226 prompt emission}
\label{table1}
\tablehead{
Sect. & Model \tablenotemark{(a)}  &
$\alpha$  & 
$\beta$ & 
E$_{p}$(keV) &
$\chi^{2}_{\nu}$ (dof) 
} 
\startdata
A & Power-law & $-$1.97 $\pm$ 0.41  &  & & 1.32 (7)   \nl
  & Band law&   $-$1.88 $\pm$ 0.21 & $<$ $-$2.0 & $<$ 4 & 1.16 (5)  \nl
 & & & &  &  \nl
B & Power-law & $-$1.66 $\pm$ 0.07 &  &  & 1.35 (7)  \nl
  & Band law &  $-$1.25 $\pm$ 0.25 & $-$2.6 $\pm$ 0.7 & 61 $\pm$ 15 & 0.91 (5)   \nl
 & & & &  &  \nl
C & Power-law & $-$2.14 $\pm$ 0.14 &   &  & 0.85 (7) \nl
  & Band law&  $-$1.76 $\pm$ 0.38 & $<$ $-$2.0 & $<$ 7 & 1.2 (5)  \nl
 & & & &  &  \nl
D & Power-law & $-$2.17 $\pm$ 0.13 &  &   & 1.4 (7)  \nl
  & Band law&  $-$1.83 $\pm$ 0.28 & $<$ $-$2.0 & $<$ 6 & 2.1 (5)  \nl
 & & & &  &  \nl
E & Power-law & $-$2.06 $\pm$ 0.16 &  &   & 1.36 (7) \nl
\enddata
\tablenotetext{(a)}{Band law refers to 
the smoothed broken power-law proposed by Band et al. (1993): $\alpha$
and $\beta$ are the power--law photon 
indices below and above the break energy E$_{0}$, respectively. 
E$_{p}$=E$_{0}$(2+$\alpha$)} 
\end{deluxetable}

\clearpage

% Figure 1
\figcaption[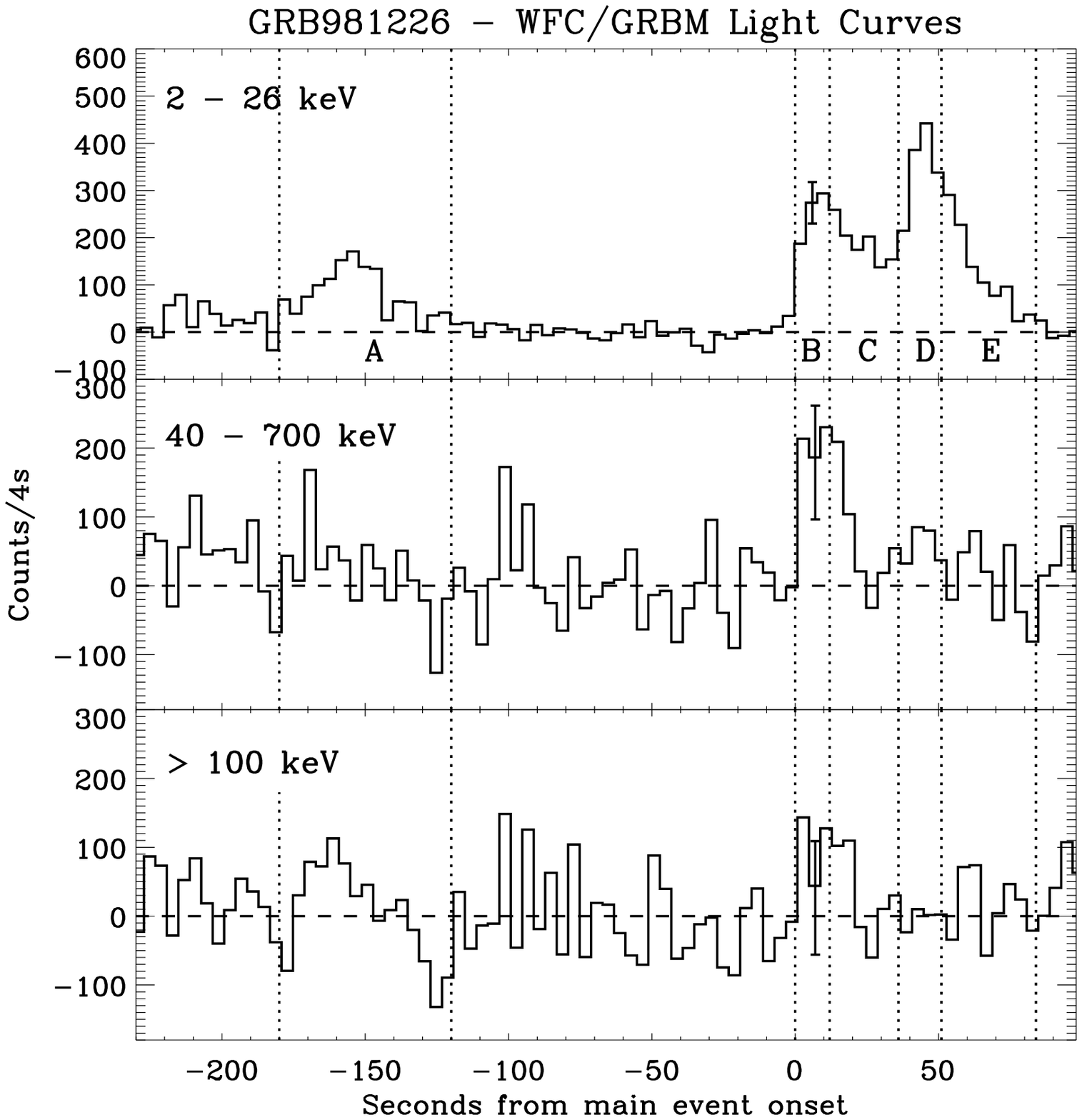]{Light curves of GRB981226 in three energy bands,
after background subtraction (see text). The zero abscissa corresponds to 1998 Dec. 26,
09:47:25 UT. The time slices on which the spectral analysis was performed 
are indicated by vertical dashed lines. The X-ray precursor start time corresponds to
$-185$~s.}

% Figure 2
\figcaption[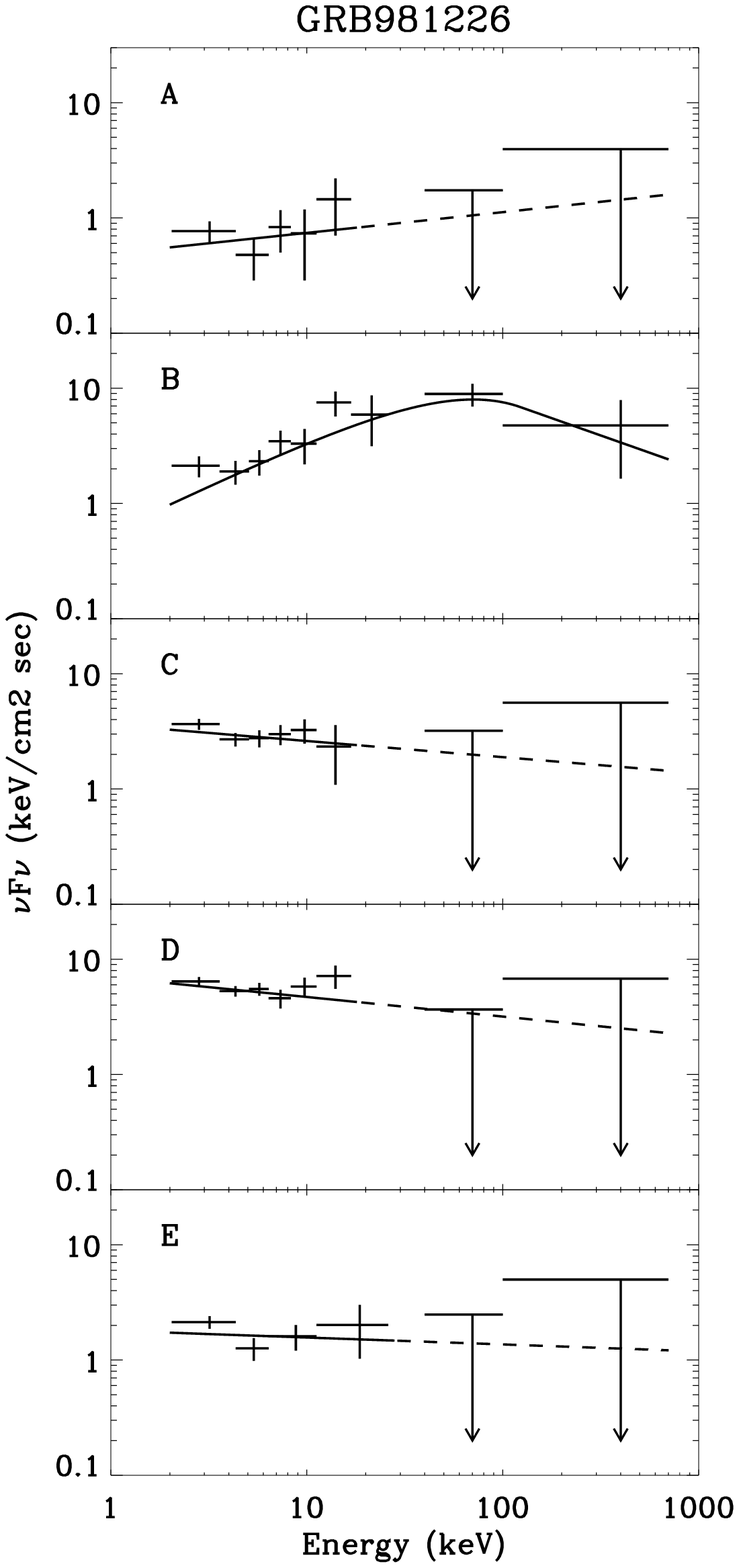]{$\nu F_\nu$ spectrum of the 5 time slices in which we
divided the burst time profile ($\nu$ is the photon
energy in keV and $F_\nu$ is the specific energy flux in
keV~cm$^{-2}$~s$^{-1}$~keV$^{-1}$). Continuous curves for slices A, C, D and E 
represent the power-law  fits to WFC data, while that for slice B represents the
fit  with a Band law \cite{Band93} to  WFC+GRBM data (see also Table~1). The dashed lines 
represent the extrapolations of the spectral fits to higher energies.}

% Figure 3
\figcaption[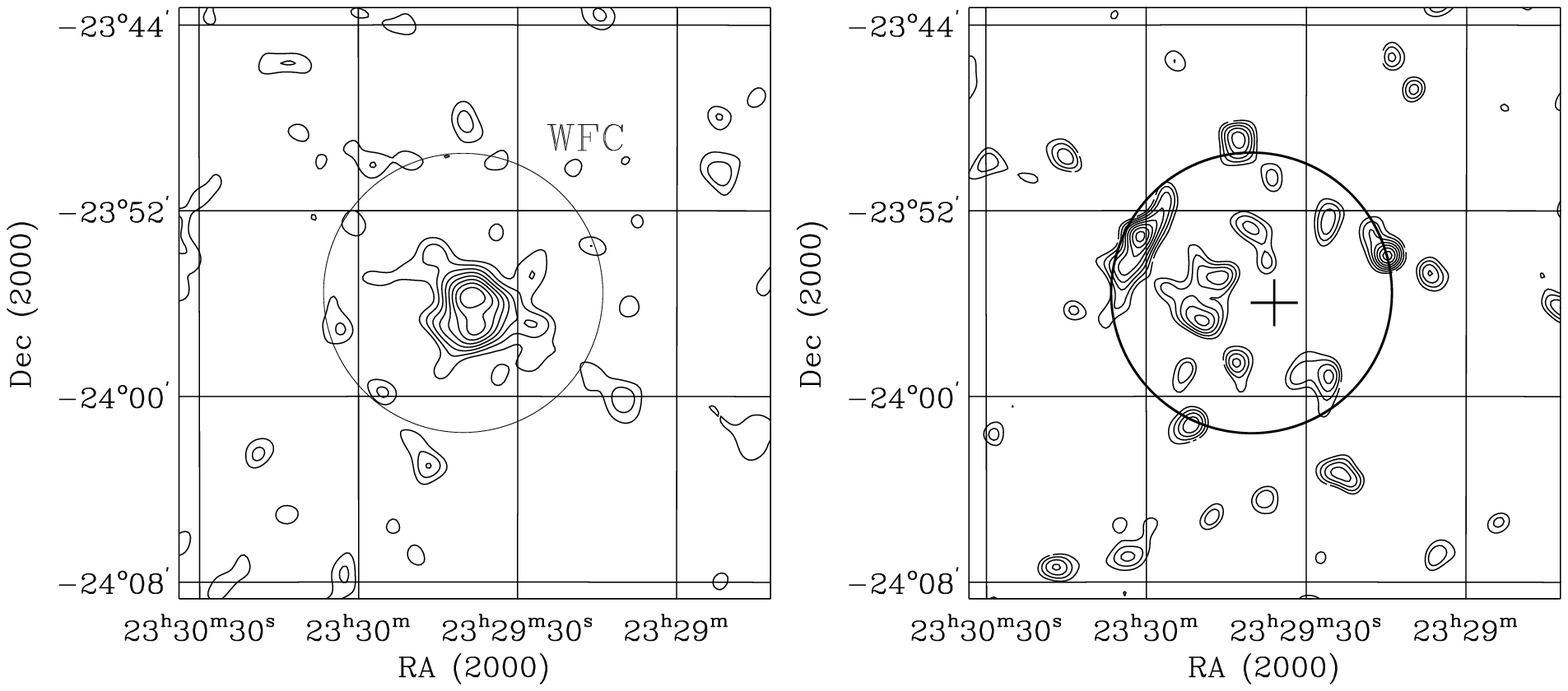]{False colour MECS images (2-10~keV) of the field
of GRB981226 obtained during TOO1 (left) and TOO~2 (right). }

%Figure 4
\figcaption[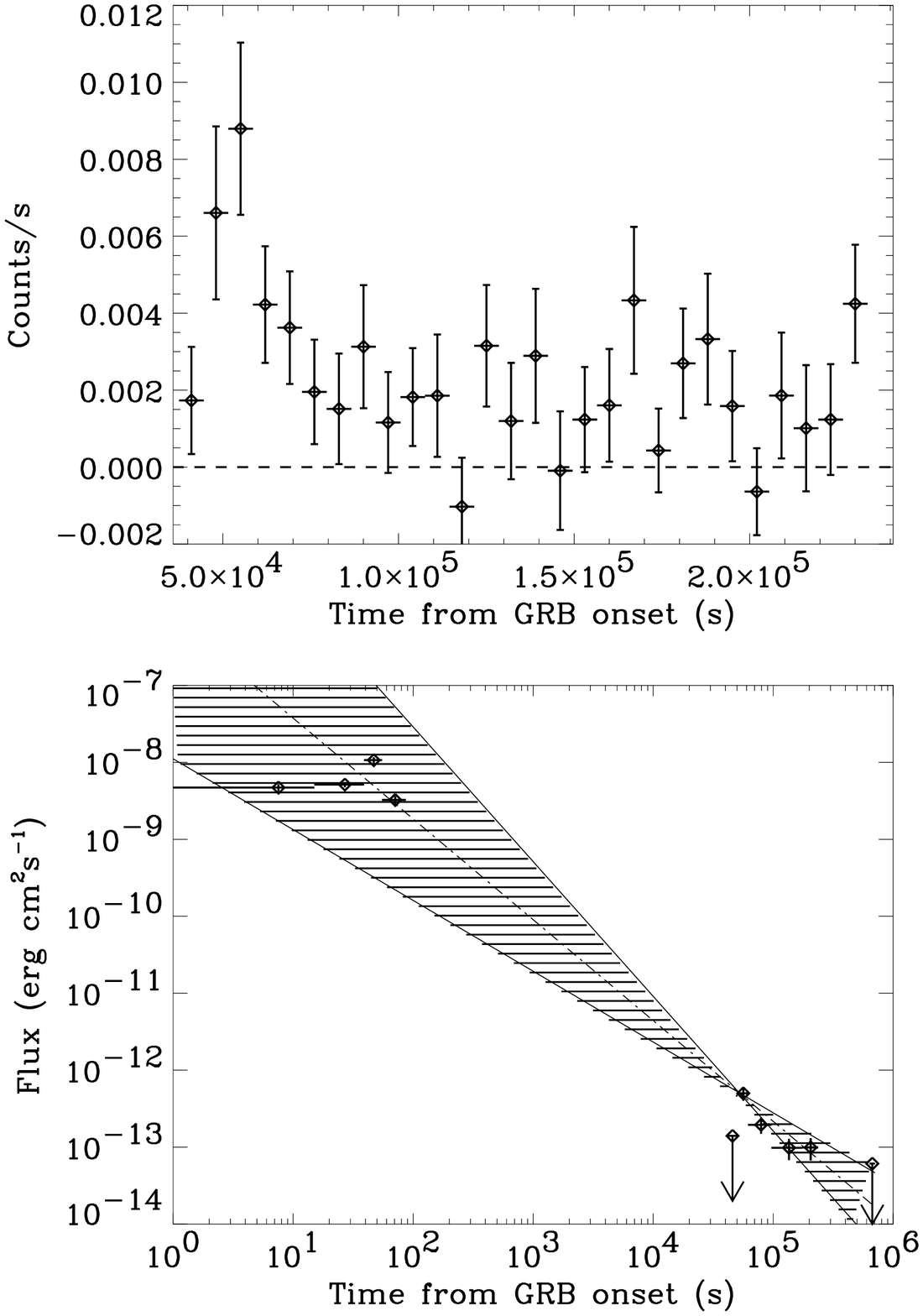] {{\it Top:} 2-10 keV MECS light curve of 
1SAX J2329.6-2356 in time bins of 7000~s each. Vertical bars represent  
$\pm$1~$\sigma$ errors.\\
{\it Bottom:} 2-10 keV WFC and MECS data. Superposed to the data points is the 
power law curve (dashed line) obtained from the best fit to the late afterglow data, 
except the first  bin with the upper limit. Also shown is the uncertainty region in the 
slope (1.31) derived from the best fit at 90\% confidence level. 
}

%\fi

\clearpage

\begin{figure}
\label{fig:timeprofile}
\plotone{fig1.ps}
\ifPreprintMode
\vspace{2cm}

\fi
\end{figure}

\clearpage

\begin{figure}
\label{fig:power}
\epsscale{0.5}
\plotone{fig2.ps}
\ifPreprintMode
\vspace{2cm}

\fi
\end{figure}

\clearpage

\begin{figure}
\label{fig:mecs_image_1_2}
\epsscale{1.0}
\vspace{-1cm}
\plotone{fig3.ps}
\ifPreprintMode

\fi
\end{figure}

\clearpage

\begin{figure}
\label{fig:lc}
\epsscale{0.9}
\plotone{fig4.ps}
\ifPreprintMode

\fi
\end{figure}

\end{document}